What do family caregivers of Alzheimer's disease patients desire in smart home technologies? Contrasted results of a wide survey.

V. RIALLE (1), C. OLLIVET (2), C. GUIGUI (3), C. HERVÉ (4)

(1) Senior lecturer, Laboratory TIMC-IMAG UMR CNRS 5525, University Joseph Fourier & Grenoble University Hospital, France.
(2) President of French association France-Alzheimer Seine-Saint Denis, Le Raincy, France ; and President of CODIFA (Association for the coordination of families of Alzheimer's disease patients in the Île-de-France region).
(3) Student in biostatistics, adviser at the laboratory TIMC-IMAG.
(4) Laboratory of Medical Ethics, Faculty of Medicine, University René Descartes, Paris, France.

**Summary**

**Objectives** – The authors' aim was to investigate the representations, wishes, and fears of family caregivers (FCs) regarding 14 innovative technologies (IT) for care aiding and burden alleviation, given the severe physical and psychological stress induced by dementia care, and the very slow uptake of these technologies in our society.

**Methods** - A cluster sample survey based on a self-administered questionnaire was carried out on data collected from 270 families of patients with Alzheimer's disease or related disorders, located in the greater Paris area. Multiple Correspondence Analysis was used in addition to usual statistical tests to identify homogenous FCs clusters concerning the appreciation or rejection of the considered technologies.

**Results** - Two opposite clusters were clearly defined: FCs in favor of a substantial use of technology, and those rather or totally hostile. Furthermore the distributions of almost all the answers of appreciations were U shaped. Significant relations were demonstrated between IT appreciation and FC's family or gender statuses (e.g., female abstractFCs appreciated more than male FCs a tracking device for quick recovering of wandering patients: p=0.0025, N=195).

**Conclusions** - The study provides further evidence of the contrasted perception of technology in dementia care at home, and suggests the development of public debates based on rigorous assessment of practices and a strict ethical aim to protect against misuse.

# I. Introduction

This article presents a survey on the views and perceptions of innovative smart home technologies (SHT) by family caregivers (FCs) (including any voluntary caregiver) of patients with Alzheimer's disease or related disorders (AD), as well as what these FCs expect of these SHT.

Alzheimer's disease is not curable and strongly linked to ageing, and most AD patients live at home with the permanent help of a FC (mostly husband/wife or children). The illness features memory loss, and a slow disintegration of personality and physical control, with manifestation of aggressiveness, wandering, incontinence, disinhibition, binge-eating, hallucinations, delusions, and depression (1-3). As the illness develops, patients become totally dependant on their caregivers, who in turn may develop more or less severe physical exhaustion, negative thoughts, a depressive state, and are threatened by a higher mortality risk (4, 5).

This stark reality is strongly contrasted by recent progress in smart homes technologies including communication and assistive technologies, and innovative telecare supports for elderly and cognitively impaired people (6-10). Such progress is developing especially in the domain of wandering and fall detection, household robotics and domotics, user friendly videoconferencing communication for homecare and social connectedness, thus giving rise to innovative community telehealth network programs (7, 10-15).

Today, despite markedly available new devices and the growing development of such programs, family caregivers remain physically and psychologically overstrained by the heavy care burden, which has a bearing on the quality of care they give. Distribution and uptaking of these new tools and resources is so poorly developed that personalized care plans and coping aids do not even mention them (16). So, there is a need to better understand the hows and whys of the discrepancy between the available technology and the way AD FCs perceive it. Studies on needs, perceptions, and expectations of AD patients and their caregivers regarding assistive technologies have been studied so far through three main approaches: a) direct users' position through end user focus groups or in-depth face-to-face interviews; b) self-administered questionnaire; and c) ethnographic studies.

End user focus groups or in-depth face-to-face interviews are aimed at collecting the users' opinion through free expression, individually or in small groups. They are widely used by searchers or evaluators for the assessment of specific tools and services (6, 17-20). They are usually costly and time consuming, and thus usable only for small scale assessment procedures.

The self-administered questionnaire approach is based on a set of questions, preferably very simple for the sake of understanding, sent to the target persons with a prepaid envelope for returning the filled questionnaire, or filled out over the telephone. This method is used in medium and large-scale studies for statistical purpose. It has been successfully applied to technology in the AD domain (19, 21).

The ethnographic study method consists in the in-depth understanding of people and context through direct observation of living situation by researchers. It was recently applied to the AD domain (19, 22, 23). This method can only be applied to a small number of participants and a small-scale environment, thus raising two issues: the representativeness of the observed subjects, and the influence of the presence of the observer on observed people or situation. It is not rare to meet a mix of methodologies in extended studies (19).

One of the most thorough technological needs analyses was performed by K.Z. Haigh *et al.* (19). They identified a list of nearly 300 technological appliances which would be of interest to elders, their caregivers (formal or informal), and other interested parties (*e.g.* insurance); opportunities of use were classified into general categories such as communications, activity monitoring, user monitoring, environment monitoring, reasoning, memory support, workload support, social support, event detection, and others. Then, by means of a Decision Matrix and Six Sigma (6σ) analyses, they determined the importance factor associated with each assistance need based on prevalence, contribution to institutionalization, impact on care giving resources, and limitations on elder functional ability. The ranked list of assistance needs based on an importance factor score (in parentheses) is the following: Medical monitoring (9.0), Medication management (7.5), Mobility (7.5), Caregiver burnout (7.5), Dementia (6.0), Eating (6.0), Toileting (6.0), Safety (6.0), Isolation (5.5), Transportation (5.5), Housekeeping (3.5), Money management (3.5), Shopping (3.5), Wandering (3.5), Usability (3.0), Equipment use (2.0), Hallucinations (2.0), Alcohol use (1.5), Pressure sores (1.0).

Our study for a better understanding of FCs' perception of technology was named ALICE (for 'Alzheimer, Information, Communication and Ethics') and was performed on a large sample of French FCs of AD patients living in the greater Paris area. Medical gravity and social impact of AD in the French population has been thoroughly studied in recent years. For instance, the REAL.FR study was a prospective longitudinal study conducted by the French network on Alzheimer's disease (24), and the PAQUID survey was a prospective population-based cohort study on normal and pathological aging after 65 years which included around 1,500 elders aged 75 years and older (25). Furthermore Thomas *et al.* (26) conducted the PIXEL study, the goal of which was to demonstrate the parameters influencing French caregivers' quality of life, and their possible link with AD patients' quality of life. The main features of the ALICE study sample are homogeneous with these earlier studies. For instance with regard to PIXEL: patients age range (PIXEL: $80.2 \pm 6.8$ [61-97]; ALICE: $78 \pm 8.1$ [54-98]); caregivers age range (PIXEL: $65.7 \pm 12.8$ [33-92]; ALICE: $64 \pm 12.9$ [31-92]); caregivers sex distribution (PIXEL: 63% women, 37% men; ALICE: 65.8% women, 34.2% men).

The following is a presentation of a specific part of the ALICE study: the cluster analysis of FCs' opinion based on answers to questions concerning 14 specific devices constituting a wide and representative range of SHT.

## II. Materials and Methods

### II.1 Population

The following three departments of the greater Paris area were selected for the study: Seine et Marne (Dpt # 77), Hauts de Seine (Dpt # 92), and Seine Saint Denis (Dpt # 93). They represent a large part of the Paris region including a number of satellite cities and rural areas. They were selected for their representativeness of the French social and cultural diversity: the first one is an urban and rural zone with scattered dwellings among crops; the second one has a population of well-off people and numerous head offices of big firms; the third one has a working population with a high percentage of immigrants. Each of these departments has its local Alzheimer's family association which provided the study with its own list of AD family member addresses for a total of 1,458 families.

**II.2 Methods**

*Questionnaire*

A self-administered questionnaire was written out, tested for reliability and validity, and sent by surface mail to 1,458 families of the selected population. It was intended to collect FCs' representations, wishes, and fears regarding a wide range of SHT, along with patients' abilities regarding commonly used devices. It was made-up of 50 questions divided in three main sections (Table 1) (the whole questionnaire is available on the ALICE website: http://www-timc.imag.fr/Vincent.Rialle/ALICE/Questionnaire_ALICE.pdf).

|    | Section | Topics | Questions # |
|----|---------|--------|-------------|
| A) | General information | Gender, demographics, lifestyle, and health status | 1-18, 33, 46 |
| B) | Technology | (a) Current skills and practice | 19-20, 31 |
|    |            | (b) Viewpoint on specific technologies | 21-24, 26-28, 29a, 29b, 30, 34-35, 39, 41 |
|    |            | (c) Wishes, fears, meaning of life, and technology | 25, 32, 36-38, 40, 42-45 |
| C) | Economics | Economical effort and support | 47-50 |

Table 1: Composition of the questionnaire

Most of the questions were multi-choice. Only two of them were of the open-end type for free opinion expression. Besides usual and quite simple questions regarding the appreciation or rejection of the presented technologies, we tried to probe deeper on issues regarding the patient's intimacy and dignity, and the caregiver's questioning on the meaning of life, suffering, responsibility, and death. We tried to understand whether technology had any influence on these difficult questions. Most of the questions were composed of several elementary sub-questions. For instance question n° 18 corresponded to the 'mini-Zarit' score (27), which is composed of 7 yes/no sub-questions. The only person entitled to fill up the questionnaire was the family caregiver. Most of the questions relating to smart home technology were twofold: a first part was devoted to the caregiver's personal view, while a second part was devoted to the patient's view collected by the caregiver when possible. Because of data collection bias the latter were used only for a qualitative estimation of this view.

*Subset of questions used for the cluster analysis*

The subset of questions used for the cluster analysis is the "Viewpoint on specific technologies" (VST) part of the questionnaire (Tab. 1, section B-b). This subset was devoted to assess the FC's viewpoint on 14 specific SHT (Tab.2): fall sensor (3 questions: sewn in a belt or garment, stuck on the skin, inserted inside the body), oral call identification (1 q.), video surveillance (2q.: indoor, outdoor), tracking device (1q.), patient's assessment of functional abilities (2q.: displacement and transfer), activity of daily living (ADL) identification (1q.), cooking (2q.: oral and written advices), robot care (1q.), videoconferencing (1q.). We purposely did not include 'domotic' technologies such as remote door or window openers, since they are already widespread. In the same way, we did not try to be exhaustive with examples of innovative technologies since their number is currently growing considerably: the goal was to present an adequately varied and typical panel so as to reliably assess the caregiver's opinion.

Every question was of the multi-choice type and was structured as follows. First, it presented a short description of the technology followed by the question: "Would this technology be helpful to you?"), second, an agreement scale - 'not at all', 'little', 'moderately', 'very much' - as the set of possible answer choices.

### II.3 Statistical methods used

Beside usual tests of independence for statistical relationship analyses (Chi Square, t-test, Anova), the study used the Multiple Correspondence Analysis (MCA) (28) to identify FC clusters. Each VST question corresponds to a VST variable of the MCA.

## III. Results

### III.1 Population

350 questionnaires were sent back by FCs, among which 270 (18.5% of the 1,458 questionnaires mailed to the families) met the completion criteria for the statistical analysis. Patients were 78.3 years old (± 8.1), range 54-98 years; 67.2% were women and 32.8% men. FCs were 64 years old (± 13), range 31-92 years. 65.8% were women and 34.2% men; a lot of them were very old; 13.9% were between 80 and 89 years of age (Fig. 1). 48% of FCs was the patient's husband or wife, 43.5% was a sibling, 2.6% was a relative, and 5.9% was other. The sex ratio of FCs (0.52) and patients (0.48) were comparable. 54.6% FCs had been the patient's caregiver for over 4 years, among whom 12.8% for over 8 years. 21% did not have any free time for themselves during the day and 11% had only one hour per day. 69.4% of the patients lived in a private home (individual house or apartment), 21.3% in a nursing home, a few in sheltered housing (2.2%), and 7.1% in specialized institutions or other. 52.3% of the patients had been diagnosed with Alzheimer's disease only two years after the onset of symptoms. 73.2% FCs considered their care recipient as very dependent (rate ≥ 5 on a 0-7 scale), and 38.8% as totally dependent (rate = 7 on a 0 to 7 scale).

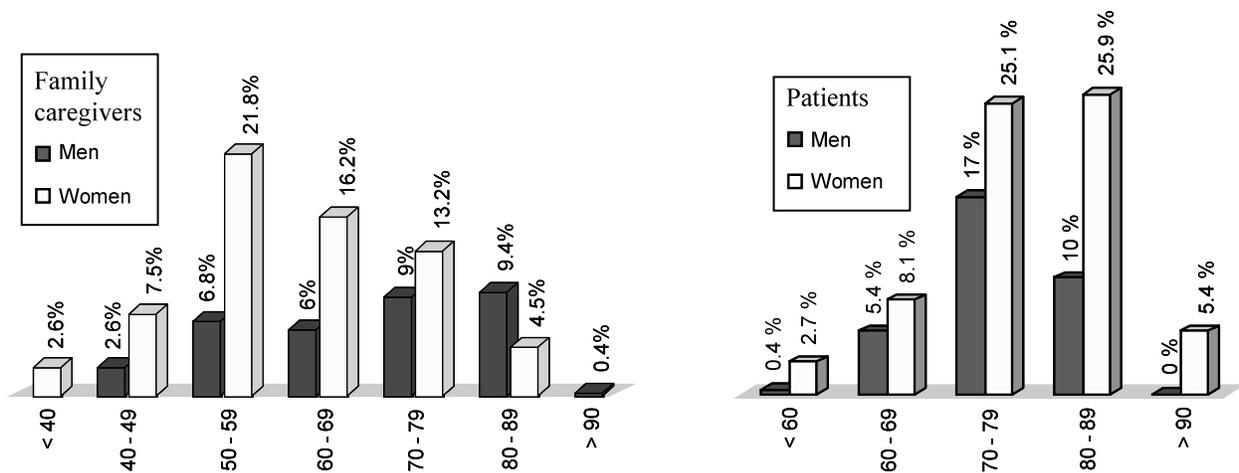

**Figure 1: Caregivers' (left) and patients' (right) age distribution**

### III.2 Technology

Distributions of answers to VST questions (Tab. 2, and Fig. 2) can be characterized as follows: 11 questions among the 14 feature a U shaped distribution of answers, peaks of answers are on the extreme items "not at all" and "very much". For the three other questions (*i.e.*, 23, 34, 35), the peaks are on "not at all" and "moderately", which flattens the U-shape. The means distribution is also U shaped.

The most appreciated technology, the tracking device (Q. 28), collected the greatest number of favorable responses (Tab. 2), with female caregivers appreciating it significantly more than male ($p=0.0025$, N=195) and younger caregivers more than older ($p<0.001$, N=196). The second most appreciated technology was the videoconferencing device for social connectedness (Q. 41) (Tab. 2), with no significant difference between male and female caregivers, or between younger and older ones. The least rejected device was the personal pocket videoconferencing device for mobile private remote surveillance of the patient (Q. 26) (Tab. 2), with also no significant difference between male and female caregivers, or between younger and older ones.

Concerning little appreciated technologies, a device for providing oral cooking advice (Q. 34) was given simultaneously the min. of 'very much' (6.4%) and the max. of 'not at all' (75.7%), and was closely followed by the fall sensor inserted inside the body under the skin ('not at all': 74.5%, 'very much': 10.8%).

| # | Question | Technology | not at all | little | Moderately | very much |
|---|---|---|---|---|---|---|
| | | | | | Percentage | |
| 1 | Q. 21 | Fall sensor stitched inside a garment or a belt | 28.9 | 12.3 | 16.6 | 42.2 |
| 2 | Q. 22 | Fall sensor pasted on the skin | 35.3 | 11.7 | 17.8 | 35.2 |
| 3 | Q. 23 | Fall sensor inserted inside the body under the skin | 74.5 | 3 | 11.7 | 10.8 |
| 4 | Q. 24 | Automatic vocal call recognition for help | 31.9 | 11.6 | 13.8 | 42.7 |
| 5 | Q. 26 | Personal pocket videoconferencing device for mobile private remote surveillance of the patient | 25 | 11 | 20.6 | 43.4 |
| 6 | Q. 27 | Video surveillance by a remote Call Centre | 34.2 | 14.3 | 15.6 | 35.9 |
| 7 | Q. 28 | Tracking device for a rapid assistance in case of wandering or running away | 30.4 | 5.1 | 11.2 | 53.3 |
| 8 | Q. 29a | Device used to assess the patient's capacity to move from one place to another one at home | 45 | 15.3 | 12.4 | 27.3 |
| 9 | Q. 29b | Device used to assess the patient's capacity to move from bed to armchair | 46.6 | 13.6 | 15 | 24.8 |
| 10 | Q. 30 | Device used for ADL identification and functional assessment | 35.1 | 12.3 | 17.5 | 35.1 |
| 11 | Q. 34 | Device providing oral cooking advice | 75.7 | 6.9 | 11 | 6.4 |
| 12 | Q. 35 | Device providing written cooking advice on a screen | 70.6 | 7.8 | 12.7 | 8.9 |
| 13 | Q. 39 | Pet robot devoted to diverting and/or lessening the patient's anxiety or agitation | 55.1 | 12.8 | 14 | 18.1 |
| 14 | Q. 41 | Videoconferencing device for social connectedness | 27.8 | 7.3 | 19.5 | 45.4 |
| | | Mean | 44.0 | 10.4 | 15.0 | 30.7 |

Table 2: Percentages of answers to the 14 VST questions "Would this technology be helpful to you?"

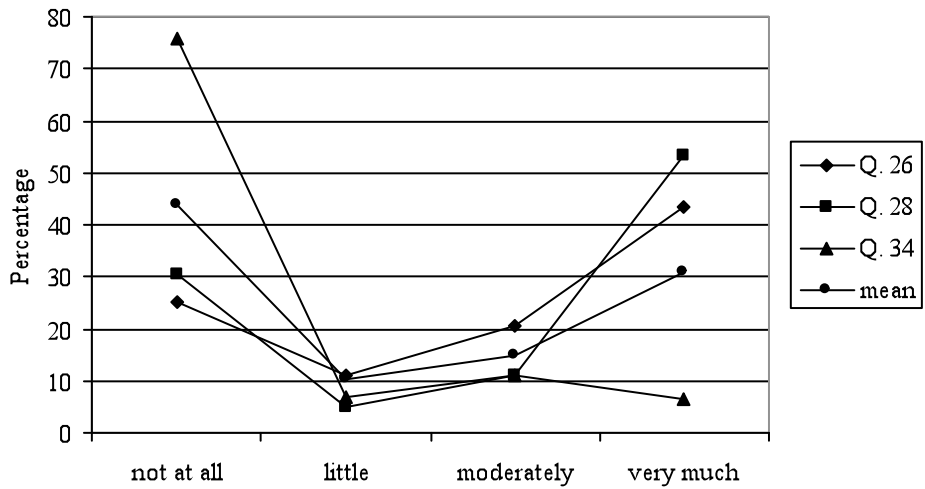

Figure 2: Mean distribution of answers to the 14 VST questions "Would this technology be helpful to you?" along with three extreme answer distributions: personal pocket videoconferencing (Q. 26), tracking device (Q. 28), and device providing oral cooking advice (Q. 34)

MCA was used to determine the various FC classes suggested by the U shaped distribution of answers. The question was: were the respondents who selected 'not at all' and respectively 'very much' globally the same for the whole set of VST questions? If this was the case, they would make up a distinctive cluster of FCs. The first MCA factorial map (Fig. 3) proves that this is indeed the case: the "not at all" items and respectively 'very much' items are clearly grouped in homogenous clusters, and the two other clusters ('little' and 'moderately') can be easily localized even though they share a few borderline answerers.

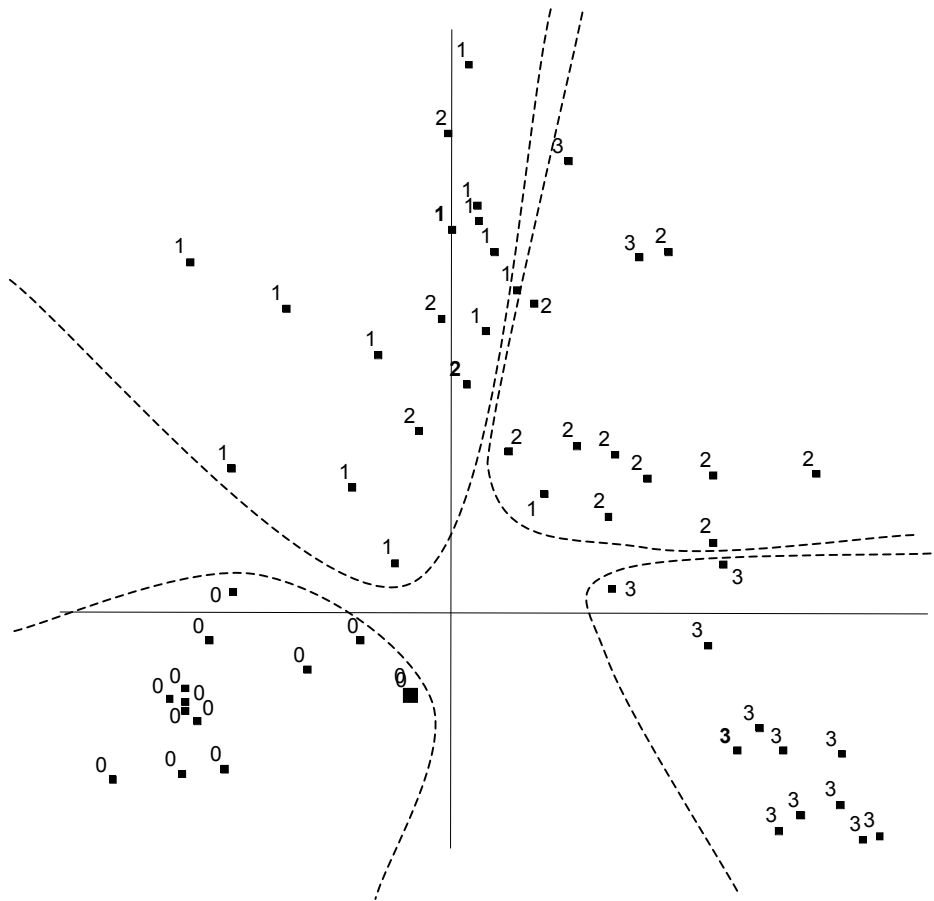

Figure 3: Clustering of caregivers (0 ≡ not at all, 1 ≡ little, 2 ≡ moderately, 3 ≡ very much). The map shows the distribution of the answers to the 14 VST variables (N=90, questionnaires with a missing answer to at least one VST question were ignored, 23.47% of the total variance is "explained" by the map)

Thus, two major opposite trends emerged from the collected data: the FCs who were mainly in favor of a substantial usage of technology for care aiding, and those who were rather or totally hostile to such a use. The tracking device was representative of such a bimodal distribution of opinions; the variance analysis resulted in a significant relation between the caregiver's age and the appreciation of the system ($p<0.0001$, N=214; mean of 'very much'= 60.7 y.o., of 'not at all'= 69.4 y.o.). This was also true for advanced devices such as company robots; sibling caregivers were more interested by these robots than the patient's husband or wife ($p=0.005$ ; N=243; mean of 'very much'= 57 y.o., of 'not at all'= 65 y.o.).

## IV. Discussion

### IV.1 Concerning the results

The results reinforce the conclusion reported by Poulson *et al* (29) according to whom there is no such things as an 'average user' in the area of assistive technologies. The fact that the two extreme items of VST questions ('not at all' and 'very much') were significantly more often selected than the medium ones ('little' and 'moderately') clearly suggests that family or voluntary caregivers are sensitive to technology. This closely matches the field reality for several technologies. For instance, the most appreciated one – tracking device (Q. 28) – is meant to answer to one of the most worrying issue of dementia care giving. According to Koester (30), when a person with Alzheimer's disease or related disorders is lost for more than 12 hours, there is a 50% chance of finding him/her injured or dead. Confronted with such a threat, caregivers must be constantly watchful, knowing that at any moment their cared-for is likely to wander or run away. Furthermore, the tracking device and service have long been awaited for by caregivers, despite ethical issues (31).

The results also stress the importance of the videoconferencing device, both for social connectedness (second most appreciated, Q. 41) and for caring (least rejected, Q. 26) with no significant difference between male and female caregivers, or between younger and older ones, in both questions. The possibility to see, hear, and talk to the care recipient at a distance was considered as highly helpful.

### IV.2 Concerning the methodology

Our study relies on an unusually large number of caregiver opinions obtained via a rather long self-administered questionnaire. Although a substantial amount of statistics was used to determine the quantitative features of the gathered data, the true intent of the study was not a statistical one. Statistics were used only to synthesize the voluminous collected information and to reveal the tendencies and remarkable relations from a statistical point of view, not to gain new academic knowledge on family caregivers. The results are those of a specific though widespread area – the greater Paris area – since it has a fairly diversified population, both culturally and economically. Moreover, these results represent only the opinions of the respondents and may not be generalizable to the entire group of FCs studied.

Due to the self-administered form of the enquiry and the age of caregivers, the first drawback of the methodology was the foreseeable difficulty for caregivers to properly understand the whole set of questions and to answer by the questionnaire themselves. The great number of questions and sub-questions increased this difficulty. The number of completely filled questionnaires (N=270, response rate = 18.5% of the 1,458 questionnaires mailed, as above mentioned) that were received in two months – without any letter of reminder – reinforce our conviction that the theme of the study was meaningful for caregivers (the total number of received answers was 351, among which 71 did not meet our filling criteria).

The second drawback of the self-administered questionnaire is the possible selection bias of the study. Indeed, it is conceivable that the people who were the most sensitive to the use of such technologies, either positively or negatively, were those most motivated to respond to the survey; these strong opinions then would appear as the result of the survey. These opinions might explain the U shaped distributions of responses. As a consequence, such a highly significant distribution shape might not be representative of the entire population. However, the hypothesis of such a bias is strongly weakened by the high rate of response obtained without any letter of reminder (18.5%), which is considered as more than expected (being usually subjectively expected around 10%). This response rate tends to prove that the inquiry was well accepted, and thus that every opinion tendency was correctly reflected in the received responses. 212 answers were received (the analysis of which would be too long for this article), which globally reflects a remarkable reflection, if not wisdom, on the part of family caregivers regarding technology. When expressed, their anger or resentment was most commonly directed against the poor consideration of the rest of the society for their burden of care and conditions of life.

## V. Conclusion

Our study determined two major opposite trends in the opinions of family or voluntary caregivers concerning smart home technology, and provided further evidence of the contrasted status of technology in caregivers' perception: rejection or mitigated appreciation of several technologies, and great confidence in the helpfulness of a few of these.

The study clearly shows that the more appreciated smart home technologies are those which increase the patient's safety while decreasing the caregiver's fear of wandering away or accident, and those which increase the caregiver's social connectedness and freedom to leave home at any time. Design of smart home technologies should take this into account to gain user acceptance. However further research will be needed to demonstrate the efficacy of technological resources in improving affect, coping, psychological well-being, and stress management, and to pass this data to family caregivers and health professionals. Such research should be carried out with an 'ethical intention' both to protect patients and caregivers against misuse (32), and to maximize beneficial effects.

**Acknowledgements**

Our heartfelt thanks to the presidents and voluntary members of the three French departmental Alzheimer's disease associations (France Alzheimer 77, France Alzheimer 92, France Alzheimer 93; affiliated to the French national Association France Alzheimer) for the taskforce they devoted to the study. A special thanks also to Evelyne Janeau, of the Grenoble Joseph Fourier University, for her useful advice on data analysis tools.

**Funding**

The data collection and a part of the statistical analysis of the survey were funded by the Rotary Club of Le Raincy, France. The rest of the statistical and cluster analyses were funded by the Conseil Régional Rhône-Alpes.